\documentclass[graybox, envcountchap]{svmult}

\usepackage{mathptmx}        
\usepackage{amsmath}
\usepackage{amssymb}
\usepackage{color}
\usepackage{helvet}          
\usepackage{courier}         
\usepackage{dirtree}

\usepackage{makeidx}        
\usepackage{graphicx}        
\usepackage{subfig}

\usepackage{multicol}        
\usepackage[bottom]{footmisc}

\usepackage{hyperref}        
\hypersetup{colorlinks=true,urlcolor=blue}

\usepackage[numbers,sort&compress]{natbib}

\usepackage[misc]{ifsym}
\usepackage{xspace}

\newcommand{\Lya}{Lyman-$\alpha$\xspace}

\newcommand{\Lyaforest}{\Lya\ forest\xspace}

\newcommand{\poned}{\ensuremath{P_\mathrm{1D}}\xspace}

\newcommand{\xithreed}{\ensuremath{\xi_\mathrm{3D}}\xspace}

\makeindex             

\begin{document}


\title{Machine-Learning Techniques for Astrophysics and Cosmology: Lyman-$\alpha$ forest}
\titlerunning{ML techniques for Lyman-$\alpha$ forest studies}
\author{Jon\'as Chaves-Montero}
\institute{Jon\'as Chaves-Montero (\Letter) \at 
Institut de F\'{i}sica d’Altes Energies (IFAE), The Barcelona Institute of Science and Technology, Edifici Cn, Campus UAB, 08193, Bellaterra (Barcelona), Spain \\ \email{jchaves@ifae.es}}


\maketitle

\abstract{The Lyman-$\alpha$ forest refers to the series of absorption features observed in the spectra of distant quasars that are produced by neutral hydrogen in the intergalactic medium. Observed over a wide range of redshifts with both ground- and space-based facilities, the Lyman-$\alpha$ forest provides a powerful probe of numerous physical processes, including the thermal state of intergalactic gas, the timing and topology of cosmic reionization, the expansion history of the Universe, the growth of cosmic structure, massive neutrinos, and the nature of dark matter. This chapter reviews the transformative impact of machine-learning techniques on Lyman-$\alpha$ forest analyses, particularly in overcoming the computational and methodological limitations of traditional approaches. We discuss a broad range of machine-learning applications, including the automated characterization of individual absorption systems, improved reconstruction of the intrinsic quasar continuum, accelerated emulation of hydrodynamical simulations, and the development of simulation-based analyses, field-level inference methods, and three-dimensional reconstruction techniques for the underlying matter density field. As current and upcoming surveys continue to increase both the volume and precision of \Lyaforest observations, ML-driven pipelines are becoming an essential component of next-generation astrophysical and cosmological analyses.}

\section{Introduction}
\label{sec:intro}

The \Lyaforest consists of a multitude of absorption features observed in the spectra of distant quasars, produced by intervening neutral hydrogen along the line of sight \cite{Gunn1965, Lynds1971}. These absorptions arise in the intergalactic medium (IGM) --- the diffuse gas that fills the space between galaxies --- and are sensitive to both its ionization and thermal state. Because most baryonic (ordinary) matter resides in the IGM \cite{Fukugita1998, Shull2012, HernandezMonteagudo2015, Nicastro2018, deGraaff2019, Tanimura2019, ChavesMontero2021}, and because the IGM traces the underlying dark matter distribution on large scales \cite{MiraldaEscude1996, Muecket1996, Theuns1998, Viel2004}, the \Lyaforest provides a powerful probe of the expansion history of the Universe, the growth of cosmic structure, the nature of dark matter, the thermal evolution of the IGM, and the timing, topology, and sources of cosmic reionization. Over the past several decades, the \Lyaforest has been extensively studied both observationally and theoretically, and has become a cornerstone observable for investigations of the high-redshift Universe. Comprehensive reviews can be found in \cite{Rauch1998, Weinberg2003, Meiksin2009, McQuinn2016}.

The study and application of the \Lyaforest have been transformed by successive generations of spectroscopic surveys. At moderate spectral resolution, successive phases of the Sloan Digital Sky Survey (SDSS \cite{York2000, Eisenstein2011, Dawson2016}) provided, for the first time, statistical samples of quasar spectra large enough for precision cosmology. These datasets enabled the first detections of baryon acoustic oscillations (BAO) in the \Lyaforest \cite{Busca2013, Slosar2013, Delubac2015, Bautista2017, deSainteAgathe2019, duMasdesBourboux2020} and in its cross-correlation with quasars \cite{FontRibera2014, duMasdesBourboux2017, Blomqvist2019}, yielding precise measurements of the cosmic expansion history. The large number of spectra also enabled stringent constraints on massive neutrinos and the primordial power spectrum through precise measurements of the small-scale matter power spectrum \cite{McDonald2006, PalanqueDelabrouille2013, PalanqueDelabrouille2015, Chabanier2019, PalanqueDelabrouille2020, Fernandez2024, Walther2025}. The ongoing Dark Energy Spectroscopic Instrument (DESI \cite{DESI2022_intro, DESI2022_introa}) represents an order-of-magnitude increase in the number of observed quasar spectra, leading to substantial improvements in constraints on the expansion history \cite{DESI2025_LYABAODR1, DESI2025_LYABAODR2}, the small-scale matter power spectrum \cite{ChavesMontero2026}, and enabling measurements of the growth rate of structure \cite{Cuceu2025}.

At higher spectral resolution, ground-based instruments such as Keck/HIRES \cite{OMeara2021}, VLT/X-shooter \cite{Lopez2016}, and VLT/UVES \cite{Murphy2019} resolve absorption features down to sub-kpc scales for hundreds of quasars at intermediate redshifts \cite{OMeara2015, Irsic2017b, Murphy2019, OMeara2021}, while space-based instruments like \textit{HST}/COS \cite{Danforth2016} and JWST/NIRSpec \cite{Jakobsen2022} extend \Lyaforest studies to low and high redshift, respectively. Taken alone or in combination with moderate-resolution datasets, these observations enable detailed constraints on the thermal state of the IGM \cite{Ricotti2000, Dave2001, Bolton2008, Lidz2010, Bolton2014, Nasir2016, Boera2019, Walther2019, Gaikwad2021, Villasenor2022, Ho2025, Hu2025}, the history, topology, and sources of cosmic reionization \cite{MiraldaEscude1998, MiraldaEscude2003, Fan2006, Bolton2007, Mesinger2010, Mortlock2011, McGreer2011, Becker2013, Keating2014, Becker2015, McGreer2015, Shull2015, Khaire2015, Greig2017, Banados2018, Bosman2018, Davies2018, Eilers2018, Greig2019, Fahad2020, Khaire2019, Puchwein2019, FaucherGiguere2020, Yang2020, Wang2020, Yang2020_DW, Zhu2021, Bosman2022, Greig2022, Zhu2022, CurtisLake2023, Jin2023, Umeda2024, Kakiichi2025, Meyer2025, Mason2026, Umeda2026}, the nature of dark matter \cite{Afshordi2003, Viel2005, Seljak2006_DM, Palazzo2007, Boyarsky2009, Viel2013, Dvorkin2014, Baur2016, Baur2017, Hui2017, Irsic2017, Yeche2017, Garzilli2019, Garzilli2021, Rogers2021a, Rogers2021b, Villasenor2023, Rogers2025}, and extensions of the standard cosmological model such as interacting dark sectors \cite{Garny2018} and early dark energy scenarios \cite{Goldstein2023}. 

Machine-learning (ML) techniques are increasingly being adopted in the analysis of \Lyaforest observations, driven not only by the scale of modern datasets but also by the computational and statistical challenges involved in their processing, modeling, and interpretation. On the observational side, one major challenge is the extraction of the \Lyaforest from raw quasar spectra \cite{Oke1982, Press1993, Bernardi2003}, where subtle continuum-fitting errors can propagate into significant biases in \Lyaforest statistics \cite{FontRibera2012, Slosar2013, Blomqvist2015, Busca2025}. On the theoretical side, a central difficulty lies in connecting the complex, non-linear physics of the IGM to observable \Lya statistics, a relation that is typically calibrated using computationally expensive hydrodynamical simulations \cite{Bolton2009, Borde2014, Lukic2015, Bolton2017, Walther2019, Rossi2020, Pedersen2021, Bird2023, Doughty2023, Puchwein2023, Chabanier2024, Walther2025}. The high computational cost of these simulations makes dense exploration of the cosmological and astrophysical parameter space infeasible with conventional inference approaches. These challenges motivate the development of computationally efficient and flexible methodologies, many of which are now based on modern ML techniques.

In this chapter, we review the basic physics of the \Lyaforest and its principal applications in astrophysics and cosmology, with particular emphasis on the growing role of ML techniques throughout the full analysis pipeline, from data processing to parameter inference. We begin in Section~\ref{sec:obs} by reviewing the main observational probes of the \Lyaforest and their applications in different scientific contexts. In Sections~\ref{sec:physics} and ~\ref{sec:reduction} we introduce the physics processes governing the forest and the main applications of ML techniques on data reduction. In Section~\ref{sec:simulating}, we describe the principal methods used to simulate the \Lyaforest, while in Section~\ref{sec:inference} we detail different strategies employed for inference. Finally, in Section~\ref{sec:conclusions}, we summarize recent progress and discuss the future role of ML in next-generation \Lyaforest studies.
\section{Observations of the \Lyaforest}
\label{sec:obs}

\begin{figure}
    \centering
    \includegraphics[width=\linewidth]{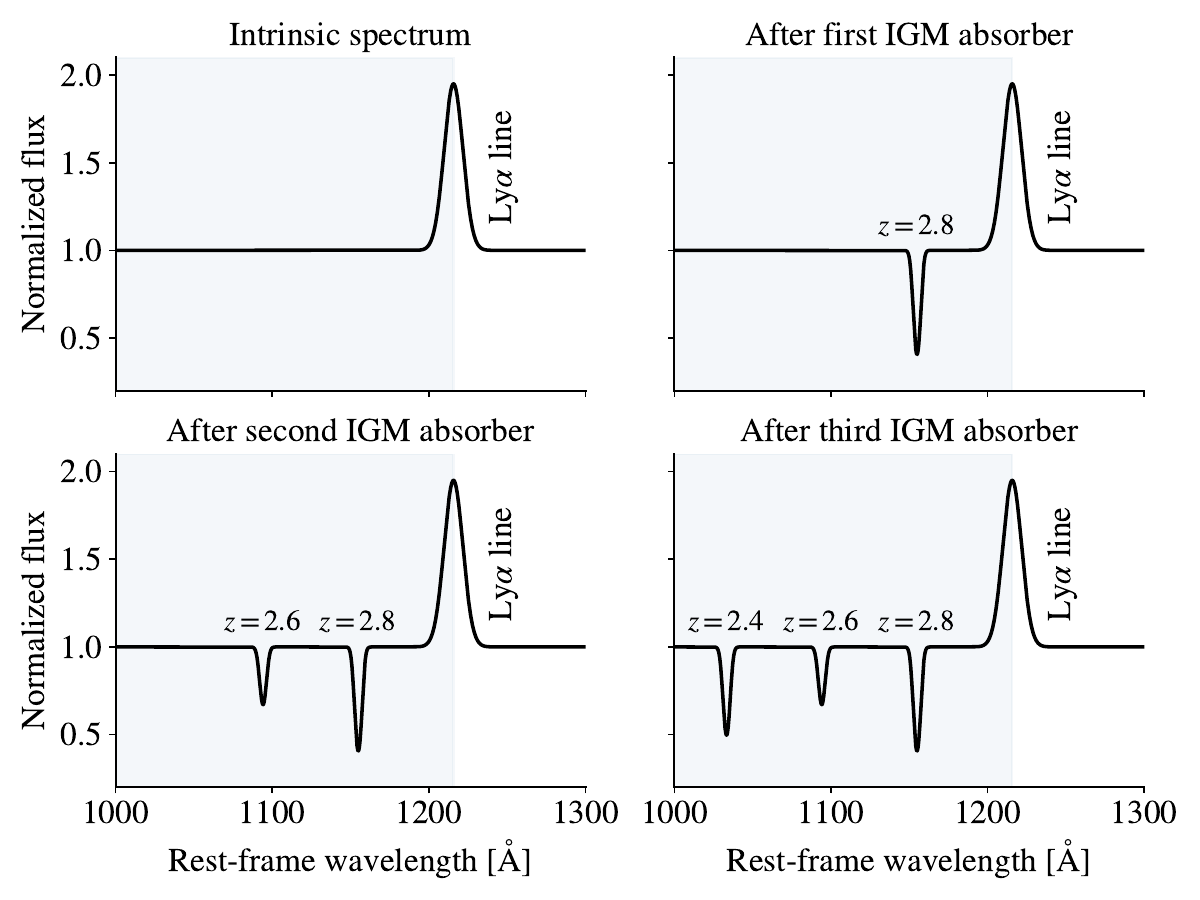}
    \caption{
    Illustration of the progressive build-up of the \Lyaforest in the quasar rest frame. The top-left panel shows the intrinsic spectrum of a quasar at $z_\mathrm{QSO}=3$. The subsequent panels show the same spectrum after successive absorption by intervening neutral hydrogen along the line of sight, corresponding to gas at redshifts $z_{\rm HI}=2.8$, 2.6, and 2.4, respectively. In the quasar rest frame, the absorption lines are produced at $\lambda_\alpha (1+z_{\rm HI})/(1+z_{\rm QSO})$.}
    \label{fig:basics}
\end{figure}

As photons emitted by a quasar propagate through the expanding Universe, their wavelengths are redshifted according to $\lambda_{\rm obs} = \lambda_{\rm rest}(1+z)$, where $\lambda_{\rm rest}$ and $\lambda_{\rm obs}$ denote the emitted and observed wavelengths, respectively, and $z$ is the redshift. Along their path, photons may encounter neutral hydrogen, and absorption occurs when the redshifted wavelength of the photons matches that of the \Lya transition ($\lambda_\alpha = 1215.67$ \AA) in the rest frame of the gas. This resonance condition can be written as $\lambda_{\rm obs} = \lambda_\alpha (1+z_{\rm HI})$, where $z_{\rm HI}$ is the redshift of the absorber. At this point, photons are resonantly absorbed and re-emitted in random directions, effectively removing them from the line of sight. Because this condition is satisfied at different redshifts along the photon trajectory, each absorption is imprinted at a distinct observed wavelength. The cumulative effect is a dense series of absorption lines blueward of the quasar’s intrinsic \Lya emission line, collectively known as the \Lyaforest \cite{Weymann1981}. We illustrate this process in Figure~\ref{fig:basics}.

\begin{figure}
    \centering
    \includegraphics[width=\linewidth]{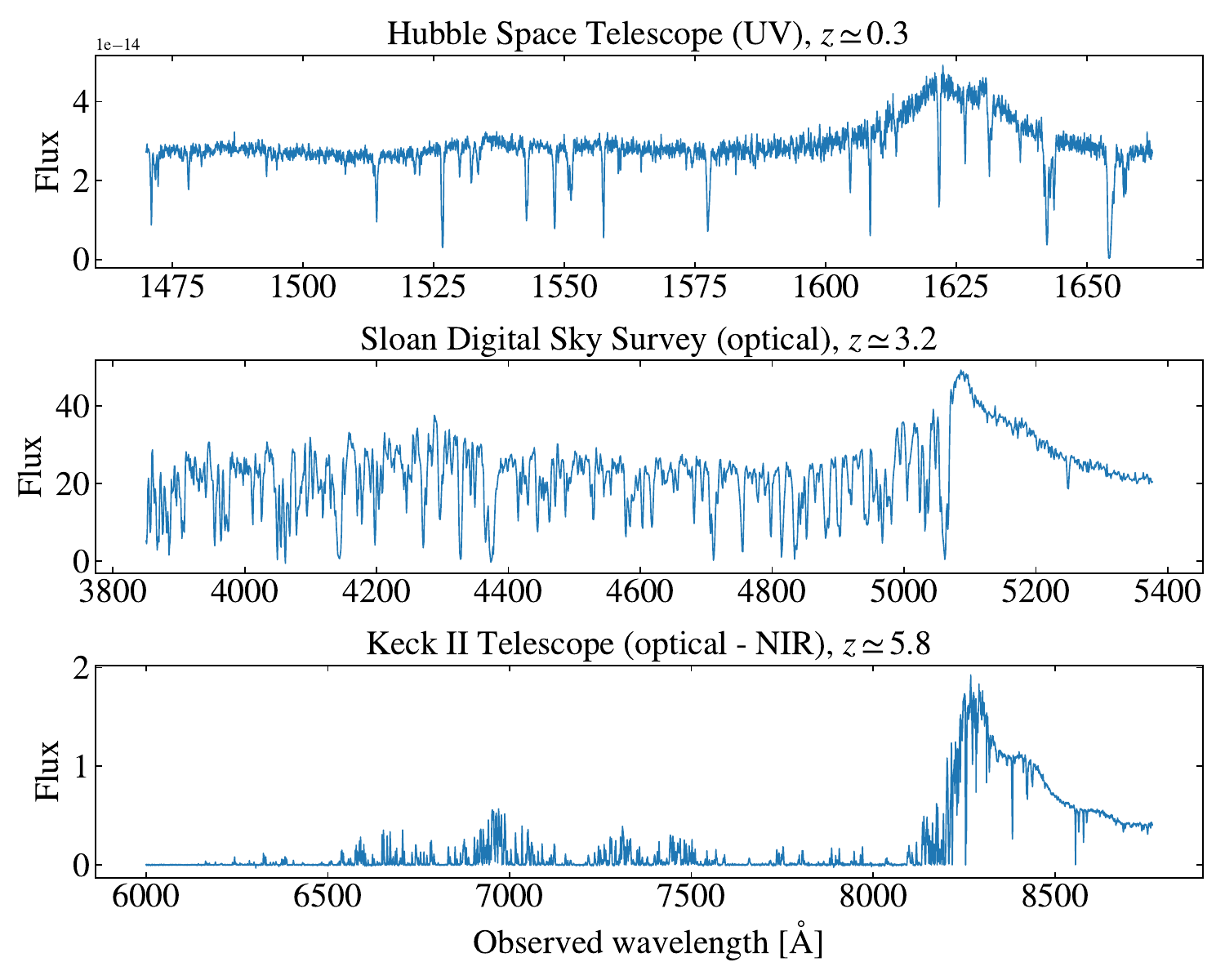}
    \caption{Observations the \Lyaforest from quasars at redshifts $z\simeq0.3$, 3.2, and 5.8 obtained with the Hubble Space Telescope, Sloan Digital Sky Survey, and the Keck II Telescope, respectively. The characteristics of the \Lyaforest evolve significantly with redshift, reflecting the progressively larger fraction of neutral hydrogen in the IGM.}
    \label{fig:quasar_spectra}
\end{figure}

We can observe the \Lyaforest in quasar spectra over a wide redshift range, with the relative contributions of cosmological information, astrophysical processes, observational systematics, and instrumental limitations varying significantly across this interval. As illustrated in Figure~\ref{fig:quasar_spectra}, both the strength and the number density of absorption features increase with redshift \cite{Sargent1980, Young1982}, reflecting the higher density of the Universe at earlier epochs and the larger fraction of neutral hydrogen in the IGM. In the following, we summarize the main applications of \Lyaforest studies across different redshift regimes.

At low redshift ($z \lesssim 2$), cosmological expansion reduces the physical gas density, and the IGM is highly ionized. As a result, the mean transmitted flux fraction increases and \Lya absorption features become sparse \cite{Meiksin2006}. The evolution is gradual, and the choice of $z \sim 2$ does not mark a physical transition, but instead reflects an observational limitation: \Lya absorption from gas at $z \lesssim 2$ falls in the ultraviolet, necessitating space-based observations since atmospheric absorption severely limits ground-based measurements \cite{Penton2004, Danforth2016}. Observations of the low-redshift \Lyaforest provide important constraints on the distribution of baryons in the Universe \citep{Shull2012, Tejos2016}, the thermal and ionization state of the intergalactic medium \cite{Ricotti2000, Dave2001, Hu2025}, and the properties of the ionizing ultraviolet background \cite{Shull2015, Khaire2015, Khaire2019, Puchwein2019, FaucherGiguere2020}.

At intermediate redshifts ($2 \lesssim z \lesssim 5$), the IGM remains highly ionized while the gas density is higher than at low redshift, giving rise to a dense \Lyaforest. From an observational perspective, this range is particularly favorable for \Lyaforest studies: the \Lya transition is redshifted into the optical, enabling observations with ground-based telescopes, and it encompasses the peak of quasar activity at $z \sim 2\text{--}3$, extending to higher redshifts where quasars remain sufficiently abundant and bright for high-quality spectroscopy \cite{Richards2006, Ross2013, PalanqueDelabrouille2016}. As a result, the \Lyaforest has been measured at moderate resolution for hundreds of thousands of quasars in successive phases of the Sloan Digital Sky Survey (SDSS \cite{York2000, Eisenstein2011, Dawson2016}) and for nearly one million quasars with the Dark Energy Spectroscopic Instrument (DESI \cite{DESI2022_intro, DESI2022_introa}). Complementary high-resolution spectra have been obtained for hundreds of quasars using instruments such as X-Shooter \cite{Lopez2016}, HIRES \cite{OMeara2021}, and UVES \cite{Murphy2019}.

Observations in this range have enabled a wide range of studies, including constraints on the thermal state of the intergalactic medium \cite{Bolton2008, Lidz2010, Rorai2013, Bolton2014, Nasir2016, Boera2019, Walther2019, Gaikwad2021, Villasenor2022, Fernandez2024, Walther2025, Ho2025, ChavesMontero2026}, and measurements of the cosmic expansion history \cite{Slosar2011, Busca2013, Slosar2013, FontRibera2014, Delubac2015, Bautista2017, duMasdesBourboux2017, deSainteAgathe2019, Blomqvist2019, duMasdesBourboux2020, DESI2025_LYABAODR1, DESI2025_LYABAODR2} and growth rate \cite{Cuceu2023, Cuceu2025}. In addition, the \Lyaforest provides constraints on the small-scale matter power spectrum \cite{Zaldarriaga2001, Croft2002, McDonald2006, Viel2006, PalanqueDelabrouille2013, Chabanier2019, Fernandez2024, Rogers2025, Walther2025, Ho2025, ChavesMontero2026}, neutrino masses \cite{Seljak2006, Bird2011, PalanqueDelabrouille2015, Yeche2017, PalanqueDelabrouille2020, Garny2021, ChavesMontero2026}, and the nature of dark matter \cite{Afshordi2003, Viel2005, Seljak2006_DM, Palazzo2007, Boyarsky2009, Viel2013, Dvorkin2014, Baur2016, Baur2017, Hui2017, Irsic2017, Garzilli2019, PalanqueDelabrouille2020, Garzilli2021, Rogers2021a, Rogers2021b, Villasenor2023, Rogers2025, ChavesMontero2026}, as well as extensions of the standard cosmological model, such as interacting dark sectors \cite{Garny2018} and early dark energy \cite{Goldstein2023}.

At high redshift ($5 \lesssim z \lesssim 6$), the \Lyaforest probes the final stages of cosmic reionization, during which hydrogen in the intergalactic medium transitions from a neutral state to the highly ionized state observed today. The neutral hydrogen fraction rises rapidly with redshift, leading to a sharp decline in the mean transmitted flux fraction and the appearance of extended regions of saturated absorption \citep{Fan2006}. As a result, spectra in this regime are no longer characterized by a continuous forest, but instead by isolated transmission spikes separated by large opaque regions \cite{Keating2014}. The statistical properties of these features provide direct insight into the history, topology, and sources of reionization \cite{MiraldaEscude2003, Bolton2007, Mesinger2010, McGreer2011, Becker2013, Becker2015, McGreer2015, Eilers2018, Bosman2018, Fahad2020, Yang2020, Zhu2021, Bosman2022, Zhu2022, Jin2023}.

At even higher redshift ($z \gtrsim 6$), the \Lyaforest becomes increasingly saturated, losing sensitivity to the ionization state of the IGM. However, constraints on the timing of reionization can still be obtained from the strength and profile of the \Lya damping-wing absorption observed around quasars \cite{MiraldaEscude1998, Mortlock2011, Greig2017, Banados2018, Davies2018, Greig2019, Wang2020, Yang2020_DW, Greig2022} and galaxies \cite{CurtisLake2023, Umeda2024, Mason2026, Umeda2026}. The observed absorption profile reflects the competition between local ionized bubbles created by galaxies and quasars \cite{Carswell1982, Bajtlik1988, Lu1991}, which allow \Lya photons to escape, and damping-wing absorption by neutral hydrogen in the surrounding IGM.
\section{The physics of \Lya absorptions}
\label{sec:physics}

The absorption of radiation by neutral hydrogen near the Lyman-$\alpha$ transition is characterized by the optical depth (e.g. \cite{Meiksin2009}),
\begin{equation}
\tau(\nu) = N_{\rm HI} \, \sigma_{\alpha}\, \phi_\mathrm{V}(\nu),
\end{equation}
where the intensity of the quasar light gets attenuated by $e^{-\tau}$, $N_{\rm HI}$ is the neutral hydrogen column density (number density of atoms per unit area), $\phi_\mathrm{V}(\nu)$ is the Voigt profile, which expresses the energy dependence of the photon cross-section, and $\sigma_{\alpha} = (\pi e^2 / m_e c)\, f_\alpha$ is the integrated cross-section over all frequencies. Here $e$ and $m_e$ are the electron charge and mass, $c$ is the speed of light, and $f_\alpha = 0.4164$ is the oscillator strength of the \Lya transition. 

The line profile $\phi_{\rm V}(\nu)$ arises from the combined effect of two physical broadening mechanisms. The first is natural (radiative) broadening, which follows from the finite lifetime of the excited state via the quantum-mechanical uncertainty principle,
\begin{equation}
\phi_{\rm N}(\nu) = \pi^{-1}\frac{A_\alpha/(4\pi)}{(\nu - \nu_\alpha)^2 + [A_\alpha/(4\pi)]^2},
\end{equation}
where $\nu_\alpha = c/\lambda_\alpha$ is the rest-frame \Lya transition frequency and $A_\alpha = 6.265 \times 10^8\,{\rm s}^{-1}$ is the Einstein coefficient of spontaneous emission. The second contribution is Doppler broadening, arising from the thermal motion of the absorbing gas,
\begin{equation}
\phi_{\rm D}(\nu) = \frac{1}{\nu_{\rm D}\sqrt{\pi}} \exp\!\left[-\frac{(\nu - \nu_\alpha)^2}{\nu_{\rm D}^2}\right],
\end{equation}
where $\nu_{\rm D} = \lambda_\alpha^{-1} b$ is the Doppler width and $b = \sqrt{2 k_{\rm B}T/m_{\rm p}}$ is the Doppler parameter, with $T$ the gas temperature, $k_{\rm B}$ Boltzmann’s constant, and $m_{\rm p}$ the proton mass. The full line profile is obtained from the convolution of these two effects, leading to the Voigt profile,
\begin{equation}
\phi_{\rm V}(\nu) = \frac{H(a,x)}{\sqrt{\pi}\,\nu_{\rm D}},
\end{equation}
where $x = (\nu - \nu_\alpha)/\nu_{\rm D}$ is the dimensionless frequency offset and $a = A_\alpha/(4\pi \nu_{\rm D})$ characterizes the relative strength of natural to Doppler broadening. The Voigt function is defined as
\begin{equation}
\label{eq:voigt}
H(a,x) = \frac{a}{\pi} \int_{-\infty}^{+\infty}
\frac{e^{-y^2}}{(x-y)^2 + a^2}\,{\rm d}y.
\end{equation}

The \Lyaforest arises primarily from absorption by mildly overdense, partially-ionized IGM gas tracing the large-scale structure of the Universe \cite{Sargent1980}. These absorbers typically have neutral hydrogen column densities $N_{\rm HI} \lesssim 10^{14}\,\mathrm{cm}^{-2}$, for which the line profile is dominated by Doppler broadening. At higher column densities, commonly associated with interstellar and circumgalactic gas, natural broadening becomes increasingly important and produces Lorentzian damping wings. These absorbers correspond to compact, mostly neutral, self-shielded systems collectively referred to as high column density (HCD) absorbers. They are commonly subdivided into Lyman-limit systems (LLSs \cite{Tytler1982}), defined by an optical depth of $\tau_\mathrm{LLS} \geq 1$, or equivalently a column density of $N_{\rm HI} \gtrsim 10^{17.2}\,\mathrm{cm}^{-2}$, and damped Ly$\alpha$ absorbers (DLAs \cite{Wolfe1986, Wolfe2005}), characterized by column densities of $N_{\rm HI} \gtrsim 10^{20.3}\,\mathrm{cm}^{-2}$. The latter threshold corresponds to the regime in which the absorber gas becomes predominantly neutral rather than ionized.

\begin{figure}
    \centering
    \includegraphics[width=\linewidth]{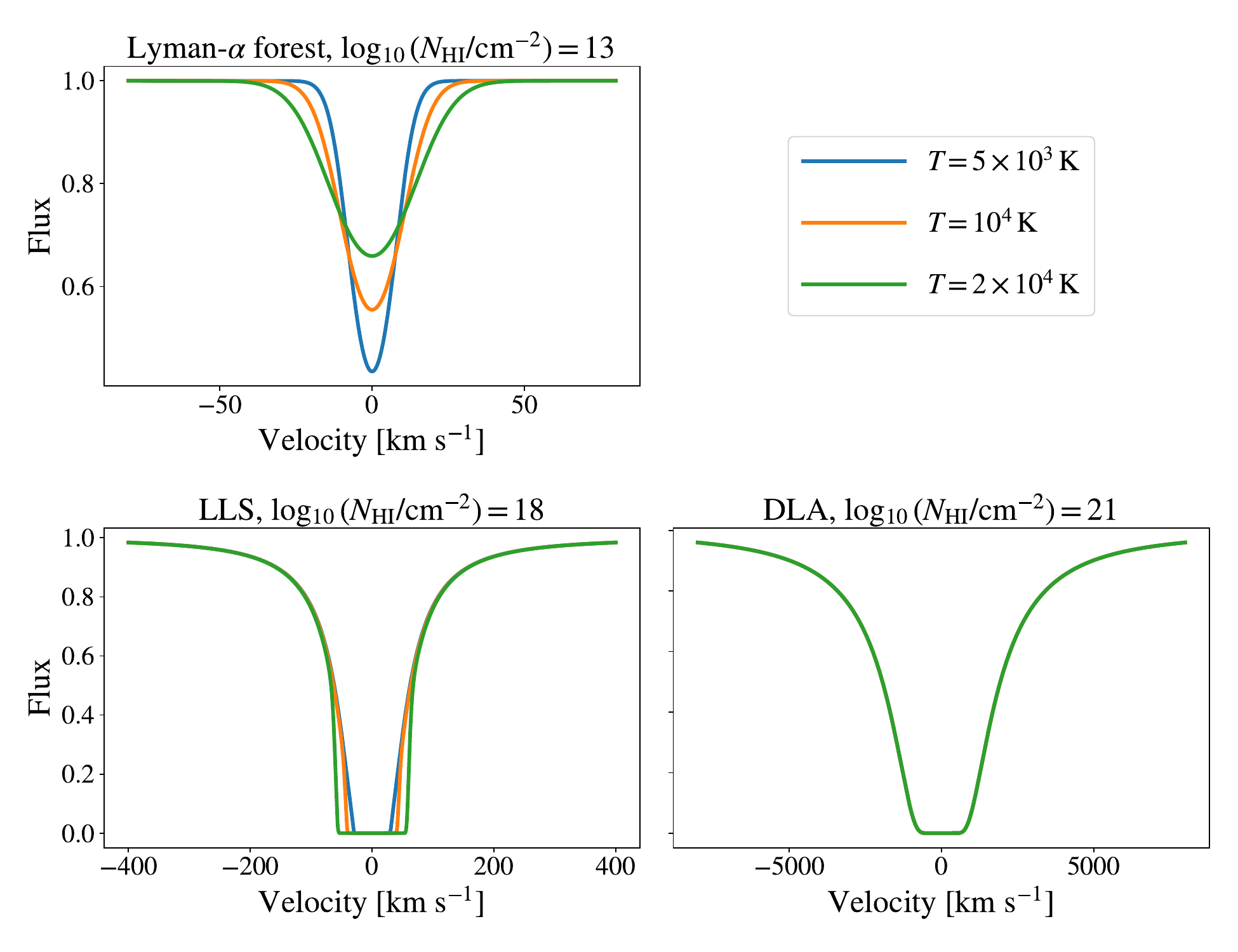}
    \caption{Dependence of the Lyman-$\alpha$ flux absorption profile on neutral hydrogen column density and gas temperature. The panels show synthetic absorption profiles for three regimes of $N_{\mathrm{HI}}$: diffuse Lyman-$\alpha$ forest gas ($N_{\mathrm{HI}} \sim 10^{13}\,\mathrm{cm}^{-2}$), Lyman limit systems ($\sim 10^{18}\,\mathrm{cm}^{-2}$), and damped Lyman-$\alpha$ systems ($\sim 10^{21}\,\mathrm{cm}^{-2}$). As the column density increases, the absorption profile becomes stronger and broader, eventually saturating and developing damping wings that progressively erase sensitivity to the gas temperature.}
    \label{fig:HCDs}
\end{figure}

Figure~\ref{fig:HCDs} illustrates the \Lya optical depth profiles produced by diffuse gas in the IGM, LLSs, and DLAs. In low column density systems, the line profile is dominated by Doppler broadening and thus retains sensitivity to the gas temperature. By contrast, in HCD systems natural broadening produces extended damping wings that become progressively less sensitive to the gas temperature as the column density increases. Because LLSs and DLAs arise in environments that are physically distinct from the diffuse IGM traced by the \Lyaforest, HCD systems are typically treated separately from the diffuse forest signal.

Finally, the \Lyaforest optical depth can be approximately related to the underlying matter density field as follows (see \cite{Meiksin2009} for a comprehensive derivation). We begin by expressing the optical depth at line center as
\begin{equation}
\tau_0 = \frac{\sigma_{\alpha}}{\nu_{\rm D}\sqrt{\pi}}\, n_{\rm HI}\,\Delta r,
\end{equation}
where the neutral hydrogen column density is written as $N_{\rm HI}=n_{\rm HI}\Delta r$, with $n_{\rm HI}$ and $\Delta r$ denoting the neutral hydrogen number density and the physical extent of the absorber, respectively. We now introduce a sequence of standard approximations valid at intermediate redshifts. Assuming that the intergalactic medium is in photoionization equilibrium, the neutral fraction is determined by the balance between photoionization and recombination processes, $x_{\rm HI} = \alpha(T)n_e/\Gamma$, where $n_e$ is the electron number density and $\Gamma$ is the photoionization rate. In the low-density intergalactic medium, the gas evolves approximately adiabatically, giving rise to a tight temperature--density relation that can be described by a power law \cite{Hui1997},
\begin{equation}
T(z,\mathbf{x}) = T_0(z)\left[1+\delta(z,\mathbf{x})\right]^{\gamma(z)-1},
\end{equation}
where $T_0(z)$ is the temperature at mean density, $\delta(z,\mathbf{x}) = \rho(z,\mathbf{x})/\bar{\rho}(z) - 1$ is the density contrast, and $\mathbf{x}$ denotes comoving position. Combining this relation with photoionization equilibrium yields the scaling $n_{\rm HI} \propto \rho^2 T^{-0.7}$. Together, these relations lead to the fluctuating Gunn--Peterson approximation (FGPA \cite{Bi1997, Croft1998, Gnedin1998, Weinberg1998}),
\begin{equation}
\tau(z,\mathbf{x}) = \tau_0(z)\left[1 + \delta(z,\mathbf{x})\right]^{\beta(z)},
\end{equation}
where $\tau_0(z)$ is a redshift-dependent normalization primarily set by the gas temperature, photoionization rate, and baryon density, while $\beta(z)=2-0.7[\gamma(z)-1]$. Consequently, the \Lyaforest optical depth acts as a biased tracer of the underlying matter density field.
\section{Data reduction}
\label{sec:reduction}

Studies of the \Lyaforest require several processing steps to extract physical information from raw quasar spectra, including instrumental calibration, sky subtraction, noise estimation, continuum fitting, and absorption-feature modeling. Among these, two central components are continuum fitting \cite{Oke1982, Press1993, Bernardi2003}, which estimates the intrinsic unabsorbed quasar emission, and Voigt-profile decomposition \cite{Morton1972, Hu1995, Lu1996, Kim2002}, which models the observed absorption features as a superposition of individual absorbers. Together, these techniques provide a framework for interpreting the forest either as a continuous transmission field tracing the underlying IGM or as collection of discrete absorption systems. We discuss continuum fitting in Section~\ref{sec:reduction_continuum} and Voigt-profile decomposition in Section~\ref{sec:reduction_individual}.


\subsection{Continuum fitting}
\label{sec:reduction_continuum}

The main observable in moderate-resolution \Lyaforest studies is the transmitted flux fraction, $F = f_{\rm obs}(\lambda)/f_{\rm cont}(\lambda)$, which is related to the optical depth via $F = e^{-\tau}$, where $f_{\rm obs}$ is the observed spectrum and $f_{\rm cont}$ is the (unknown) intrinsic quasar continuum. Accurate continuum estimation is therefore a critical step \cite{Oke1982, Press1993, Bernardi2003}, as errors propagate multiplicatively into transmitted flux fraction and all statistics derived from it. Moreover, these uncertainties are coherent over large wavelength scales, making them particularly challenging to model and mitigate.

In the analysis of large datasets of moderate-resolution spectra from surveys such as SDSS and DESI, the standard approach to continuum estimation assumes that the quasar continuum can be decomposed into a redshift-dependent mean component and a quasar-specific correction, typically parameterized as a power law \cite{Slosar2011, RamirezPerez2024}. Within this framework, the product of the intrinsic continuum and the mean IGM transmission is modeled as a smooth, slowly varying function, while quasar-to-quasar variations are described by a small set of parameters. Although this approach is computationally efficient and generally robust, it introduces continuum-estimation errors that are correlated with the underlying density field, thereby producing correlated distortions across different lines of sight \cite{Slosar2011, FontRibera2012, Slosar2013, Blomqvist2015, Busca2025}. Furthermore, the assumed parametric forms may not fully capture the intrinsic diversity of quasar continua, leading to non-negligible systematic uncertainties.

\begin{figure}
    \centering
    \includegraphics[width=0.8\linewidth]{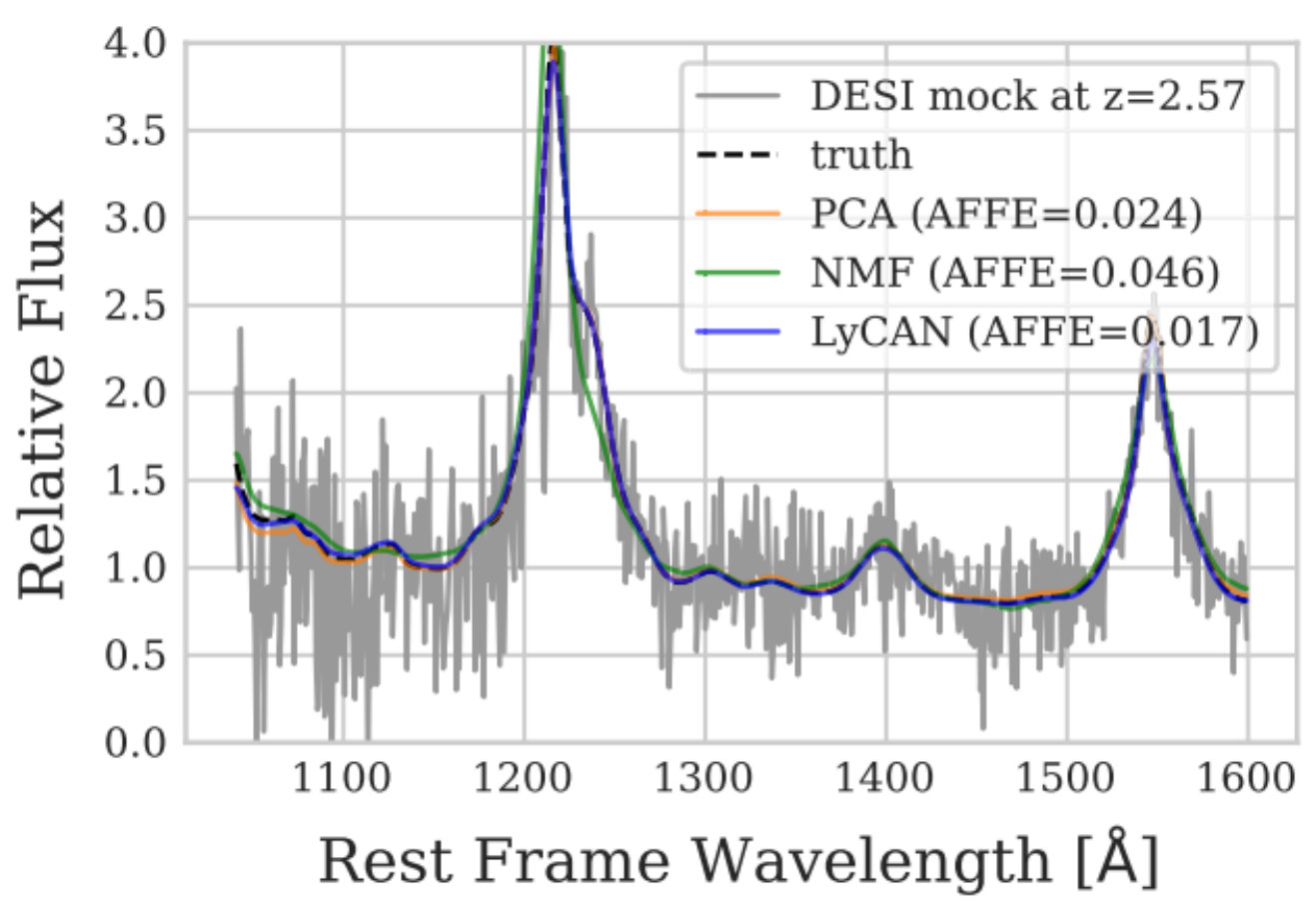}
    \caption{Reconstruction of the quasar continuum in a simulated DESI-like quasar spectrum (grey curve) using PCA (orange), non-negative matrix factorization (NMF \cite{Lee1999}; green), and convolutional neural networks (blue). The black dashed line shows the underlying continuum. The legend reports the absolute fractional flux error (AFFE), which quantifies the reconstruction accuracy. Among the methods shown, the CNN-based approach provides the most accurate recovery of the continuum. Credit: adapted from Figure~9 of \cite{Turner2024}.}
    \label{fig:continuum}
\end{figure}

To address these limitations, alternative approaches aim to infer the blue-side continuum from the red side of the quasar spectrum, where \Lyaforest absorption is absent. These include power-law extrapolation \cite{Fan2006, Yang2020} and principal component analysis (PCA \cite{Pearson1901}), with improved performance by representing quasar continua as linear combinations of basis components \cite{Suzuki2005, Suzuki2006, Paris2011, Lee2012, Davies2018}. More recently, a variety of ML extensions have been developed, including fully connected neural networks (NNs \cite{Rosenblatt1958}) \cite{Fathivavsari2020}, autoencoders \cite{Hinton2006} as in \cite{Liu2021, Pistis2025, Hahn2025}, unsupervised learning models \cite{Sun2023}, convolutional neural networks (CNNs \cite{LeCun1989}) \cite{Turner2024, Pistis2025}, and U-Net architectures \cite{Ronneberger2015, Cicek2016} as demonstrated in \cite{Pistis2025}. A comparison of several of these approaches is shown in Figure~\ref{fig:continuum}; see also \cite{Bosman2021, Pistis2025} for detailed comparisons between methods. Some of these techniques have already been applied to observational data and hold significant promise for future cosmological analyses because, unlike traditional approaches, their continuum-estimation errors are less correlated, or potentially uncorrelated, with the underlying density field.

Accurate reconstruction of the intrinsic continuum is also essential for quasar proximity-zone analyses, since continuum uncertainties must be marginalized over to constrain the timing of reionization \cite{MiraldaEscude1998, Mortlock2011, Greig2017}. Continuum inference becomes particularly challenging in this regime because the high neutral hydrogen fraction at $z \gtrsim 6$ strongly suppresses the transmitted flux blueward of the \Lya emission line, severely limiting direct access to the intrinsic quasar spectrum in the forest region. As in lower-redshift studies, PCA-based approaches \cite{Davies2018_continuum, Bosman2021} and other dimensionality-reduction techniques \cite{Hennawi2025} have been widely used to predict the intrinsic continuum from correlations with the red-side quasar spectrum. More recently, a variety of machine-learning methods have been introduced, including hybrid PCA--NN models \cite{Durovcikova2020, Durovcikova2024} and conditional normalizing flows \cite{Winkler2019, Papamakarios2019} as in \cite{Reiman2020}; see \cite{Bosman2021, Greig2024} for comparisons of different techniques.


\subsection{Voigt-profile decomposition}
\label{sec:reduction_individual}

When observed at sufficiently high spectral resolution, the \Lyaforest can be decomposed into a set of discrete absorption features. The standard approach to characterising these systems is to model each feature with a Voigt profile \cite{Morton1972, Hu1995, Lu1996, Kim2002, Prochaska2005, Prochaska2009, Rudie2012, Kim2013}, thereby inferring the neutral hydrogen column density and Doppler parameter associated with each absorber; see Equation~\ref{eq:voigt} for details. This decomposition provides a physically-motivated description of the forest, and the resulting joint distribution of $N_{\rm HI}$ and $b$ values has been extensively used to constrain the IGM temperature–density relation \citep{Schaye1999, Ricotti2000, McDonald2001}.

The decomposition of absorption features into Voigt profiles becomes increasingly challenging as the spectral resolution and signal-to-noise ratio decrease, since line blending renders the decomposition non-unique and the choice of the number of components introduces an additional degree of subjectivity. Moreover, the computational cost of traditional fitting procedures becomes substantial for large datasets, often necessitating visual inspection. Several studies have proposed automated fitting algorithms \cite{Dave1997, Carswell2014, Bainbridge2017, Gaikwad2017, Krogager2018}; however, these approaches generally do not yet achieve the same level of reliability and accuracy as human-assisted analyses.

\begin{figure}
    \centering
    \includegraphics[width=0.7\linewidth]{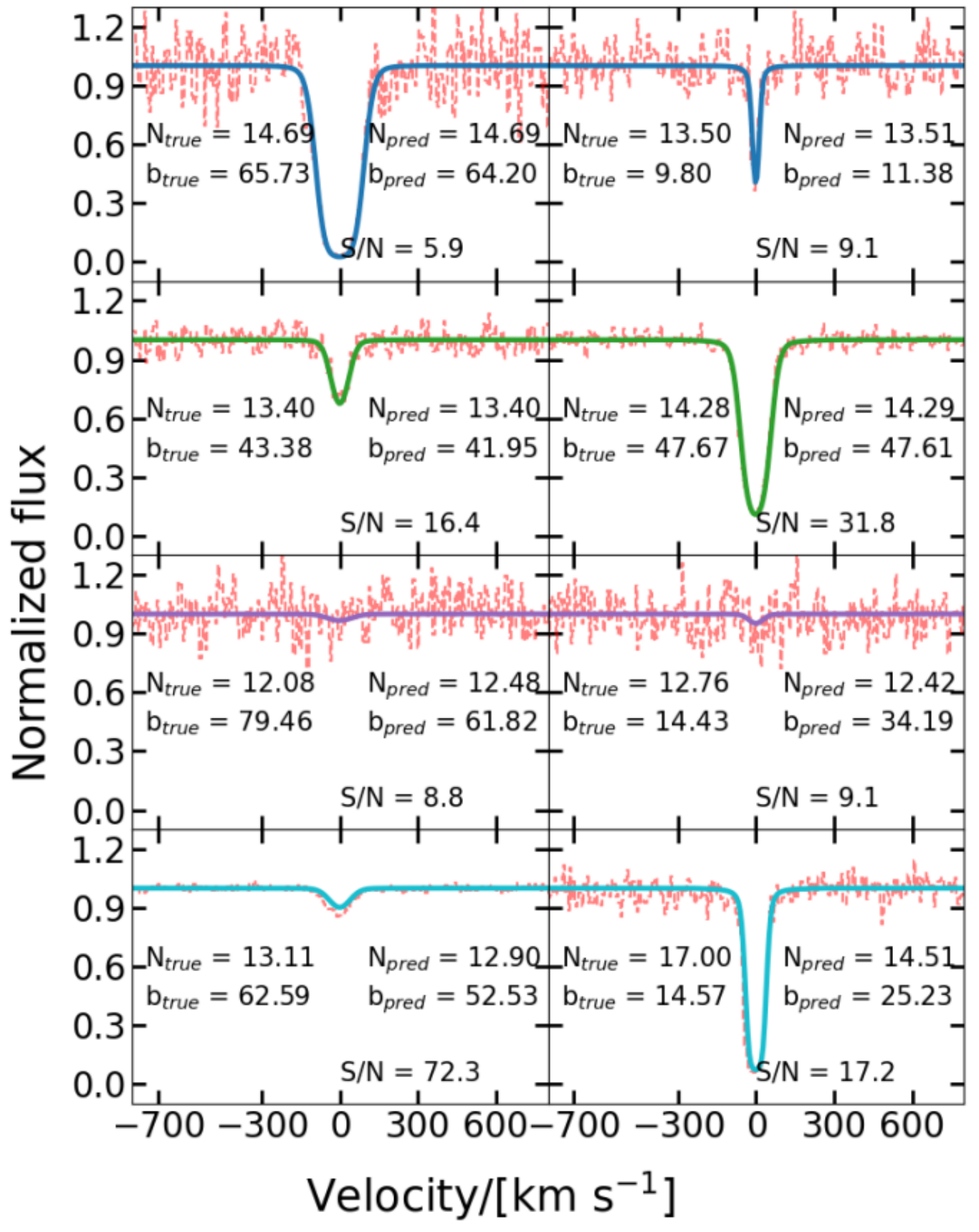}
    \caption{
    CNNs achieve high precision in identifying \Lyaforest absorption features in high-resolution spectra and in inferring the column densities and Doppler parameters of the corresponding systems \cite{Cheng2022, Jalan2024}. The figure compares the ground-truth values of these quantities for a representative set of simulated absorbers with the CNN predictions, showing excellent agreement. Credit: Figure~9 from \cite{Jalan2024}.}
    \label{fig:cnn_lines}
\end{figure}

ML methods are increasingly replacing traditional approaches because of their computational efficiency and the reduced reliance on manual inspection. CNNs have been applied to identify \Lyaforest absorbers and predict their redshifts, column densities, and Doppler parameters \cite{Cheng2022, Jalan2024} (see Figure~\ref{fig:cnn_lines}). They have also been used to identify quasars exhibiting broad absorption line (BAL) features \cite{Weymann1981, Weymann1991}, ranging from standalone CNN architectures \cite{Busca2018, Guo2019} to hybrid models combining CNNs with either residual neural networks (ResNets \cite{He2016}) \cite{Rastegarnia2022, Moradi2024} or bidirectional long short-term memory networks (Bi-LSTMs \cite{Schuster1997}) \cite{Li2025}. In addition, CNNs and Gaussian processes (GPs \cite{Rasmussen2006}) have been employed to identify and characterize DLAs, as demonstrated in \cite{Parks2018, Wang2022_DLA, Chabanier2022, Brodzeller2025} and \cite{Garnett2017, Ho2020, Ho2021}, respectively.

\section{Simulating the \Lyaforest}
\label{sec:simulating}

Simulations of structure formation solve the coupled equations of gravity, hydrodynamics, and radiative processes in an expanding Universe, thereby capturing the interplay between dark matter, baryons, and the ultraviolet background from first principles. Given a cosmological model motivated by independent observations, such hydrodynamical simulations provide predictions for the \Lyaforest that are in remarkable agreement with observations, with sufficient accuracy to discriminate between plausible variations in the parameters of the $\Lambda$CDM model and some of its extensions \cite{Cen1994, Zhang1995, MiraldaEscude1996, Hernquist1996, Rauch1997, Croft1998, McDonald2000, Meiksin2001, Croft2002, Viel2004, Viel2006, Viel2010}. Ideally, these simulations would simultaneously encompass the large cosmological volumes required for precise BAO measurements while resolving the Jeans scale of the gas, which sets the characteristic pressure-smoothing scale below which baryonic fluctuations are suppressed and determines the typical widths of \Lya absorption features \cite{Bolton2009, Lukic2015}. In practice, achieving such a dynamic range remains computationally prohibitive. As a result, current analyses rely on a combination of modeling approaches, tailored to the scales and observables of interest.

High-resolution hydrodynamical simulations remain the cornerstone for accurate modeling of the \Lyaforest \cite{Bolton2009, Lukic2015, Doughty2023, Chabanier2024}. Advances in high-performance computing have steadily increased the accessible simulation volume; however, current simulations still fall short of the volumes required for robust validation of BAO measurements. They are nevertheless ideally suited for interpreting the small-scale clustering of the \Lyaforest, which is used to constrain the thermal and ionization state of the IGM, probe the small-scale matter power spectrum, and investigate the nature of dark matter. The primary limitation of these studies is the high computational cost of hydrodynamical simulations, which prevents the dense sampling of parameter space required for accurate inference of these quantities. As a result, the standard approach relies on suites of a limited number of hydrodynamical simulations \cite{Borde2014, Bolton2017, Walther2019, Rossi2020, Pedersen2021, Bird2023, Puchwein2023, Walther2025}, combined with interpolation or ML techniques to enable parameter inference (see Section~\ref{sec:inference}).

To access the large volumes required for BAO analyses, a common approach is to generate the \Lyaforest in post-processing from gravity-only simulations --- which are substantially faster than full hydrodynamical simulations --- using the fluctuating Gunn--Peterson approximation (see Section~\ref{sec:physics}). By sacrificing some physical realism in exchange for computational efficiency, this approximation enables the rapid generation of \Lyaforest simulated observations over cosmological volumes comparable to those probed by current surveys. These mocks are widely used to test analysis pipelines and quantify the impact of systematic effects \cite{LeGoff2011, FontRibera2012, Bautista2015, Farr2020, Etourneau2024, Cuceu2025_mocks, HerreraAlcantar2025, Casas2026}, as well as for reconstruction approaches (see Section~\ref{sec:inference_field}). The mapping is commonly applied to density fields from $N$-body simulations \cite{Meiksin2001, Viel2002, Slosar2009, White2010, Lee2014, Hadzhiyska2023}, and, for even greater efficiency, to approximate gravity solvers \cite{Horowitz2019} or Gaussian and lognormal random fields \cite{LeGoff2011, FontRibera2012, Bautista2015, Farr2020, Etourneau2024}.

However, the FGPA has important limitations, including its inability to capture small-scale baryonic effects and the breakdown of the temperature--density power-law relation in high-density or shock-heated regions \cite{Lukic2015, Kooistra2022}. As a result, its accuracy deteriorates on small scales. While this limitation is not critical for BAO analyses, it is becoming increasingly important for full-shape analyses of \Lyaforest clustering. To bridge the gap between FGPA-based mocks and full hydrodynamical simulations, significant effort has recently been devoted to iterative schemes that improve the accuracy of the FGPA approximation while largely preserving computational efficiency \cite{Peirani2014, Sorini2016, Peirani2022, Sinigaglia2022, Sinigaglia2024}. 

\begin{figure}
    \centering
    \includegraphics[width=0.7\linewidth]{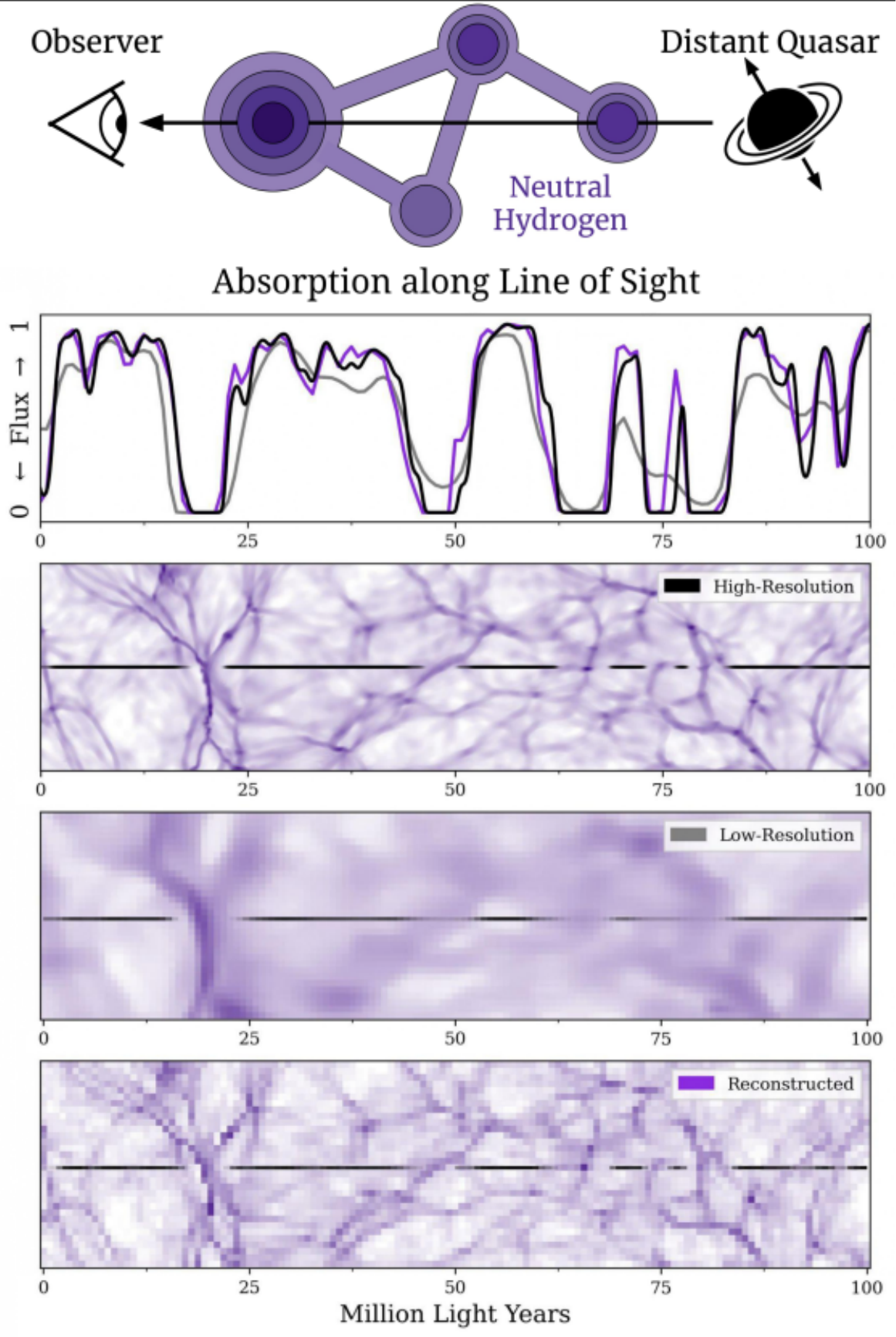}
    \caption{
    Several studies \cite{Jacobus2023, Jacobus2025, Hafezianzadeh2025} investigate super-resolution techniques in which small-scale structure in large, low-resolution hydrodynamical simulations is reconstructed using generative adversarial networks trained on small, high-resolution simulations. Credit: Figure~3 from \cite{Jacobus2025}.
    }
    \label{fig:superresolution}
\end{figure}

In parallel, ML techniques are increasingly being used both to accelerate traditional optimization methods \cite{Sinigaglia2022, Sinigaglia2024} and to improve the fidelity of \Lyaforest mocks. A variety of deep-learning approaches have shown promising performance in this context. For example, \Lyaforest mocks can be generated from gravity-only simulations using U-Net architectures \cite{Harrington2022, Boonkongkird2023}, or with variational autoencoders \cite{Kingma2013, Johnson2016} as in \cite{Horowitz2022}. Other studies have employed generative adversarial networks \cite{Goodfellow2014} to reconstruct small-scale structure in low-resolution hydrodynamical simulations, effectively emulating higher-resolution outputs \cite{Jacobus2023, Jacobus2025, Hafezianzadeh2025}, as illustrated in Figure~\ref{fig:superresolution}. In some cases, generative models have even been used to produce \Lyaforest mocks directly, without explicit simulation inputs \cite{ZamudioFernandez2019}. Despite these promising developments, however, such methods have not yet seen widespread adoption in cosmological analyses.

\section{Strategies for inference}
\label{sec:inference}

Bayesian inference provides a general framework for constraining cosmological and astrophysical models from observations of the \Lyaforest. Given a set of model parameters $\boldsymbol{\theta}$ and an observed data vector $\mathbf{x}_{\rm obs}$, Bayes theorem relates the posterior probability distribution of the parameters to the likelihood of the data and the prior information on the parameters,
\begin{equation}
\mathcal{P}(\boldsymbol{\theta}\mid \mathbf{x}_{\rm obs}) \propto
p(\boldsymbol{\theta})\,
\mathcal{L}(\mathbf{x}_{\rm obs}\mid\boldsymbol{\theta}),
\end{equation}
where $\mathcal{L}(\mathbf{x}_{\rm obs}\mid\boldsymbol{\theta})$ denotes the likelihood, $p(\boldsymbol{\theta})$ the prior distribution, and $\mathcal{P}(\boldsymbol{\theta}\mid \mathbf{x}_{\rm obs})$ the posterior distribution. Posterior sampling is typically performed using Monte Carlo techniques, enabling robust estimates of parameter degeneracies, credible intervals, and derived quantities.

\Lyaforest analyses can be broadly classified according to the modeling strategy adopted for the likelihood and the level of compression applied to the data vector. In summary-statistic approaches, the flux field is compressed into a reduced set of observables $\mathbf{s}$, such as correlation functions or power spectra, which can be efficiently compared with theoretical predictions or simulations through an explicit likelihood model, most commonly approximated as a multivariate Gaussian distribution. Within this class of analyses, BAO measurements generally rely on models based on linear theory (see Section~\ref{sec:inference_theory}), whereas analyses of the one-dimensional flux power spectrum (\poned) typically employ emulator-based frameworks (see Section~\ref{sec:inference_emu}), in which fast surrogate models trained on suites of simulations interpolate the dependence of the summary statistics on the model parameters, $\mathbf{s} = \phi(\boldsymbol{\theta})$.

\begin{figure}
    \centering
    \includegraphics[width=\linewidth]{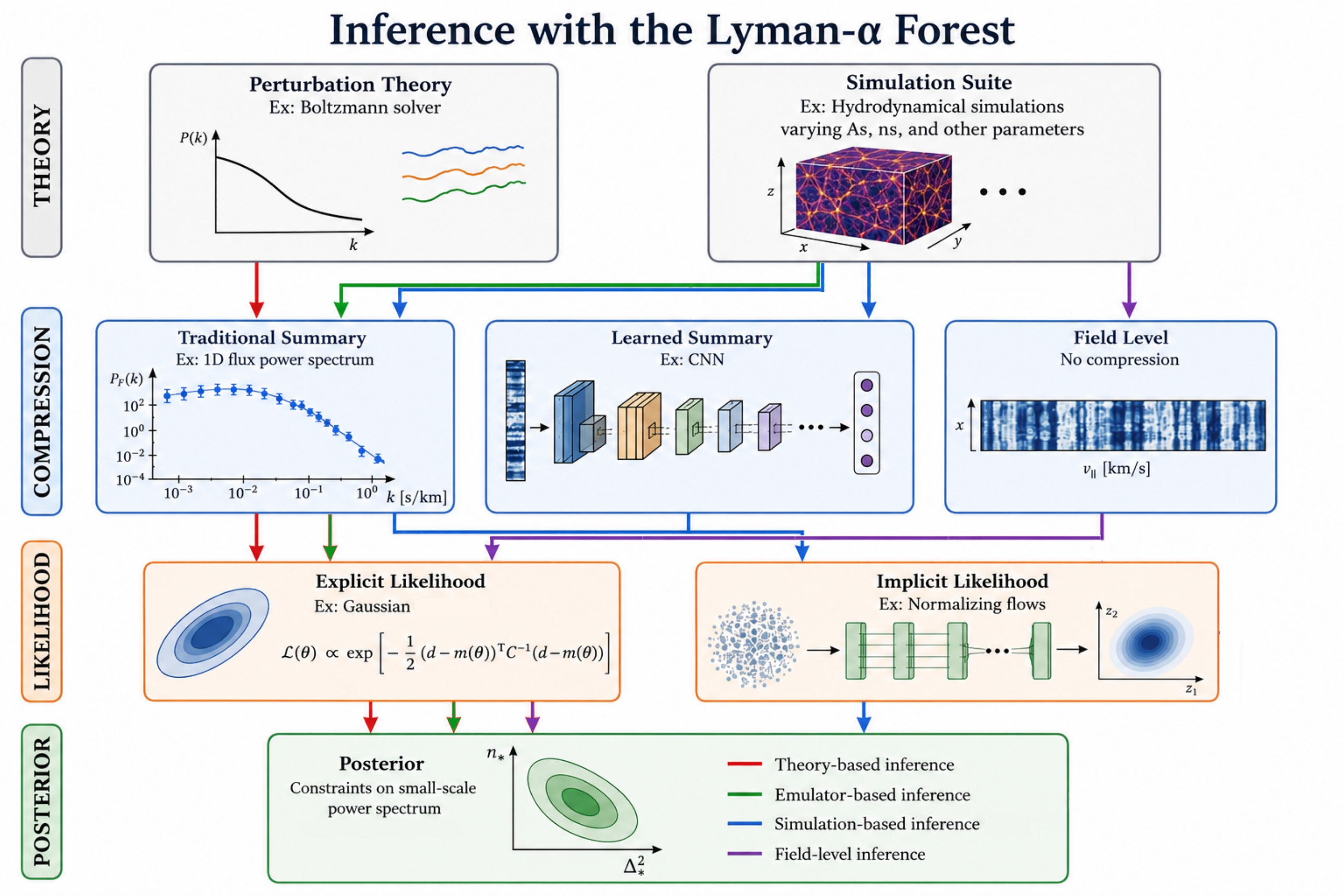}
    \caption{
    Schematic overview of inference methods employed in \Lyaforest studies. BAO and full-shape analyses typically fall within the category of theory-based inference, while most \poned analyses rely on emulator-based approaches. Simulation-based inference methods have been used to learn the relationship between Voigt-profile decompositions of the \Lyaforest and the physical properties of the intergalactic medium. Field-level inference methods, in turn, have been employed to reconstruct the underlying three-dimensional matter distribution from sparse \Lyaforest observations.}
    \label{fig:sim_inference}
\end{figure}

Simulation-based inference (SBI) avoids the need for an explicit analytic likelihood by learning either the likelihood or the posterior distribution directly from simulations, thereby enabling inference in the presence of highly non-Gaussian and non-linear data distributions (see Section~\ref{sec:inference_sbi}). Finally, field-level inference seeks to maximize the cosmological information extracted from the data by modeling the observed flux field directly, rather than compressed summary statistics, typically combining forward modeling, differentiable simulations, and high-dimensional Bayesian inference techniques (see Section~\ref{sec:inference_field}). A schematic overview of these inference approaches is shown in Figure~\ref{fig:sim_inference}.

\subsection{Theory-based inference}
\label{sec:inference_theory}

Summary statistics are widely used in \Lyaforest analyses to compress the transmitted flux from many spectra into a compact set of observables that retain much of the relevant cosmological and astrophysical information. In particular, the three-dimensional correlation function (\xithreed) measures spatial correlations of flux fluctuations as a function of separation, while \poned quantifies fluctuation along individual lines of sight as a function of scale. These statistics can be efficiently compared with theoretical models at relatively low computational cost. This approach is further motivated by the fact that, for a Gaussian random field, two-point statistics capture all information, making them a natural baseline for cosmological inference.

Different analyses of the \Lyaforest rely on distinct summary statistics, reflecting the range of physical scales they probe. BAO analyses are based on measurements of \xithreed and provide constraints on the cosmic expansion history and dark energy \cite{DESI2025_LYABAODR1, DESI2025_LYABAODR2}. Because the BAO feature lies on very large scales, these measurements can be interpreted within linear theory supplemented by the Kaiser model \cite{Kaiser1987}, which accounts for bias and large-scale redshift-space distortions, together with corrections for astrophysical and systematic effects \cite{FontRibera2012_HCDs, Bautista2017, duMasdesBourboux2017, duMasdesBourboux2020, Gordon2023, Cuceu2025_mocks, Casas2026}. Similarly, full-shape analyses of \xithreed are typically well described by linear theory with the aforementioned corrections down to quasi-linear scales ($r \sim 30$–$40$ Mpc) \cite{Cuceu2023, Cuceu2025}. 

In contrast, analyses based on \poned require a more detailed treatment, as this statistic effectively integrates over transverse modes of the three-dimensional flux power spectrum, mixing contributions from linear and nonlinear scales at each wavenumber. While there is growing interest in describing \Lyaforest clustering using the effective field theory of large-scale structure \cite{Ivanov2024}, which extends theoretical modelling into the mildly nonlinear regime \cite{deBelsunce2025, Hadzhiyska2025, Ivanov2025, deBelsunce2026, Karacayli2026}, the standard approach remains based on hydrodynamical simulations, which provide accurate predictions down to highly nonlinear scales \cite{Bolton2009, Lukic2015, Doughty2023, Chabanier2024}.


\subsection{Emulator-based inference}
\label{sec:inference_emu}

The primary limitation of hydrodynamical simulations is their computational cost, driven by the need to resolve the Jeans (pressure-smoothing) scale of the gas for accurate \Lyaforest predictions \cite{Bolton2009, Lukic2015}. This requirement limits simulation volumes below those probed by current surveys and makes dense sampling of the cosmological and astrophysical parameter space of interest infeasible. In practice, \poned analyses rely on suites of a limited number of simulations \cite{Borde2014, Bolton2017, Walther2019, Rossi2020, Pedersen2021, Bird2023, Puchwein2023, Walther2025}, combined with interpolation techniques to enable inference. Before the adoption of machine-learning methods, these included direct interpolation schemes \cite{McDonald2006, Bird2011, Irsic2017, Boera2019, Gaikwad2020, Gaikwad2021, Garzilli2021, Hooper2022, Villasenor2022, Wolfson2023, Villasenor2023} and Taylor expansions around a fiducial model \cite{Viel2006, Viel2009, PalanqueDelabrouille2013, Wang2013, Viel2013, PalanqueDelabrouille2015, Yeche2017, PalanqueDelabrouille2020}.

The first ML-based inference methods adopted in \poned analyses were emulator-based approaches \cite{Habib2007, Heitmann2009}, in which fast surrogate models --- typically GPs or neural networks --- replace direct evaluations of computationally expensive simulations. Over the past decade, the community has developed a large number of GP-based emulators \cite{Bird2019, Pedersen2021, Pedersen2023, Fernandez2022, Bird2023}, which, combined with Gaussian likelihoods, have been extensively used to constrain the properties of dark matter \cite{Rogers2021a, Rogers2021b, Rogers2022, GarciaGallego2025}, the properties of IGM gas \cite{Rorai2013, Rorai2017, Hiss2019, Walther2019}, and the small-scale matter power spectrum \cite{Fernandez2024, Walther2025, Ho2025, ChavesMontero2026}.

However, the steady improvement in observational precision --- particularly with the advent of DESI, which has increased the number of \Lyaforest spectra available for \poned analyses by a factor of several \cite{Ravoux2025, Karacaily2025} --- is progressively making emulator accuracy a dominant limiting factor \cite{ChavesMontero2026}. The accuracy of these surrogate models depends critically on the number, volume, and resolution of the simulations used for training. At the same time, the computational cost of GP-based emulators scales cubically with the number of training samples ($\mathcal{O}(N^3)$), making them increasingly difficult to apply as training datasets grow in size and complexity. These limitations are motivating the development of more scalable inference frameworks capable of efficiently exploiting the statistical power of next-generation datasets.

To alleviate these computational costs, recent work has explored adaptive sampling strategies based on Bayesian optimization \cite{Rogers2019, Walther2021}, as well as multi-fidelity approaches that combine low- and high-resolution simulations to construct more efficient GP emulators \cite{Fernandez2022, Bird2023, Fernandez2024, Ho2025}. In parallel, alternative emulation strategies based on neural networks have also been developed. These include feed-forward NNs \cite{CabayolGarcia2023, Molaro2023}, which has been applied to measure the topology of reionization \cite{Molaro2023}, conditional normalizing flows used to jointly predict the one- and three-dimensional flux power spectra \cite{ChavesMontero2025}, and autodifferentiable inference pipelines that combine NNs with Hamiltonian Monte Carlo (HMC \cite{Duane1987}) to accelerate parameter estimation and constrain the thermal and ionization state of the IGM \cite{Jin2025, GonzalezHernandez2025a}.

\begin{figure}
    \centering
    \includegraphics[width=\linewidth]{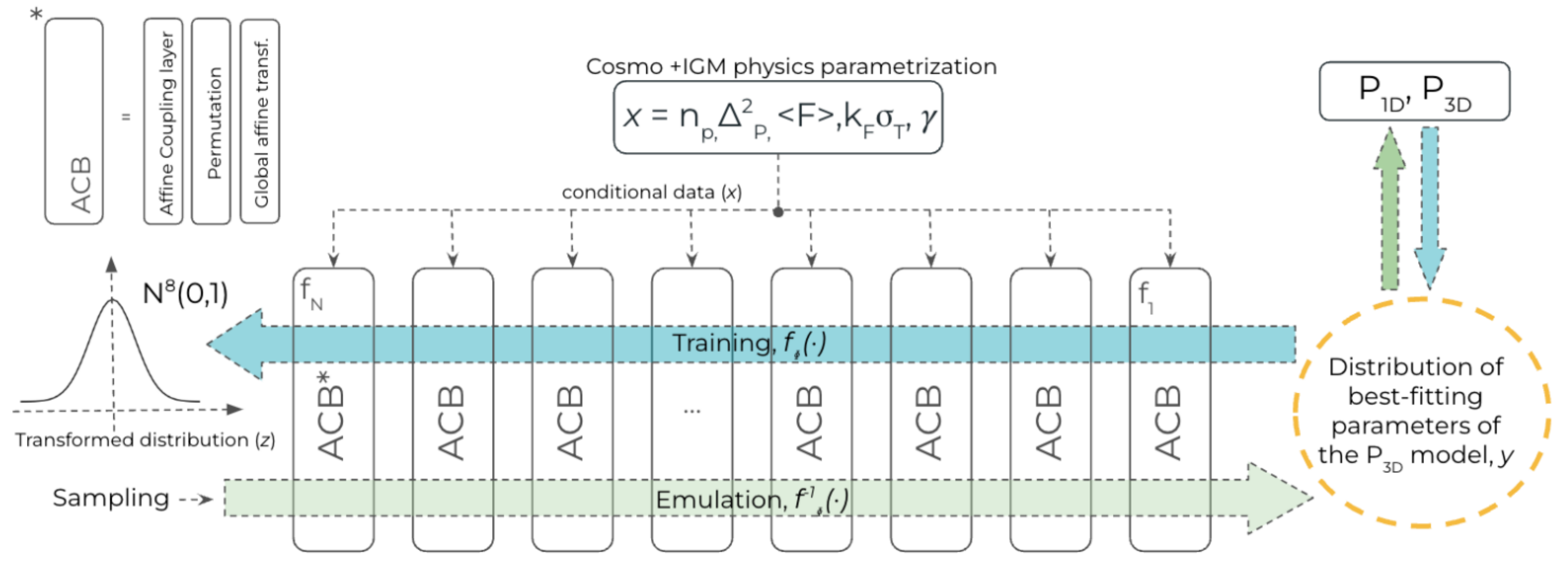}
    \caption{
    Architecture of an emulator for \Lyaforest clustering based on a conditional normalizing flow. The blue arrow denotes the training stage, during which the emulator learns a bijective mapping between summary statistics measured from a suite of hydrodynamical simulations and an eight-dimensional normal distribution. The mapping is conditioned on the cosmology and ionization and thermal state of the IGM in the simulations, and is implemented through affine coupling blocks (ACBs \cite{Dinh2016}). The green arrow indicates the emulation stage, in which the inverse mapping is applied to random samples drawn from the base distribution to generate predictions for \Lyaforest clustering. Credit: Figure~3 from \cite{ChavesMontero2025}.
    }
    \label{fig:forestflow}
\end{figure}


\subsection{Simulation-based inference}
\label{sec:inference_sbi}

Simulation-based inference (SBI \cite{Akeret2015, Papamakarios2016, Alsing2018}) provides a flexible alternative to traditional likelihood-based approaches in \Lyaforest analyses. This is because this inference approach can accommodate high-dimensional models as well as the non-Gaussian features present in the data. Instead of requiring an explicit analytic likelihood, SBI learns either the posterior distribution, the likelihood function, or the likelihood-to-evidence ratio directly from simulations, enabling efficient inference while exploiting the full information content of \Lyaforest observables. SBI has found a variety of applications in \Lyaforest studies that we summarize below.

As discussed in Section~\ref{sec:reduction}, the distribution of Doppler parameters and column densities obtained from Voigt-profile decompositions of the \Lyaforest is sensitive to both the temperature--density relation of the IGM and the UV background photoionization rate. Classical analyses typically focus on the lower cutoff of the joint distribution of these quantities \cite{Schaye1999, Ricotti2000, McDonald2001, Rudie2012, Bolton2014, Boera2014, Rorai2018, Hiss2018}, since thermal broadening imposes a minimum Doppler width at a given gas density. However, accurately determining this cutoff is challenging and susceptible to several systematic effects \cite{Rorai2018, Hiss2018}. Building on this framework, \cite{Hu2022, Hu2024, Hu2025} employed masked autoregressive flows \cite{Papamakarios2017} to learn the likelihood function and infer IGM properties from the full distribution of Doppler parameters and column densities \cite{Hiss2019}.

SBI methods have also been applied to one-dimensional \Lyaforest clustering analyses. For example, \cite{Sinigaglia2026} used masked autoregressive flows to learn the posterior distribution and constrain cosmological parameters from the one-dimensional flux power spectrum, while \cite{GonzalezHernandez2025b} employed a ResNet architecture to learn the likelihood-to-evidence ratio and characterize the epoch of reionization. Similar techniques have been used to constrain the timing of reionization and quasar lifetimes \cite{Khrykin2021} from \Lya damping-wing absorption imprinted on the proximity zones of high-redshift quasars. In this context, some studies interpolate an approximate likelihood constructed from simulations \cite{Davies2018}, whereas others approximate the posterior distribution using random forest classifiers \cite{Breiman2001}, as in \cite{Park2025}.

While many existing analyses rely on physically motivated summary statistics, recent work has increasingly explored NN-based compression schemes that can potentially recover information that is not captured by conventional summary statistics. For instance, information-maximizing neural networks have been trained to construct compressed summaries that preserve as much information as possible about cosmological parameters \cite{Charnock2018, Maitra2024}, which are subsequently used within SBI pipelines. In addition, convolutional architectures such as ResNets have been employed to extract information directly from spectral features and constrain the temperature--density relation of the IGM \cite{Nayak2024, Nayak2025}, yielding substantially tighter constraints than methods based on traditional summary statistics \cite{Chang2025}.


\subsection{Field-level inference}
\label{sec:inference_field}

Field-level inference methods operate directly on pixel-level spectra, thereby avoiding any explicit compression into summary statistics. Instead, they learn a mapping between the full transmitted-flux field and the underlying physical quantities of interest, retaining information from non-linear structure and higher-order correlations in the \Lyaforest that is typically lost in compressed analyses. Unlike SBI approaches based on learned summary statistics, these methods bypass any intermediate latent compression step. For example, CNNs trained on hydrodynamical simulations have enabled the inference of the underlying \Lya optical-depth field from noisy and saturated transmitted-flux data \cite{Huang2021}, as well as the direct inference of the thermal properties of the IGM from quasar spectra at both intermediate \cite{Wang2022, Nasir2024} and high redshift \cite{Chen2023}.

These methods have also been applied to reconstruct the underlying three-dimensional matter distribution from \Lyaforest observations along neighboring sightlines, a technique commonly referred to as tomographic reconstruction. This approach has a long history \cite{Pichon2001, Caucci2008, Cisewski2014, Lee2014, Lee2018, Ravoux2020, Newman2020, Li2021, Newman2025} and has been used to identify protoclusters \cite{Stark2015b, Lee2016, Ravoux2020}, voids \cite{Stark2015a, Krolewski2018, Ravoux2020}, the geometry of the cosmic web \cite{Caucci2008, Lee2014, Lee2016b}, and the physical properties of the IGM \cite{Muller2021}. Traditional approaches typically rely on analytical approximations and Gaussianity assumptions, such as Wiener filtering or Gaussian smoothing, to interpolate sparse sets of sightlines into a continuous three-dimensional field (see \cite{Muller2020} for a comparison of methods). Consequently, these techniques are primarily sensitive to smooth, large-scale structures in the IGM. More recent work instead employs fully differentiable forward models relating the \Lyaforest optical depth to the density, temperature, and peculiar velocity of the IGM gas, enabling the reconstruction of these physical quantities directly from \Lyaforest observations \cite{Ding2024}.

A complementary class of methods aims to reconstruct the primordial initial conditions and the corresponding matter density field directly from \Lyaforest observations, motivated by the fact that the \Lyaforest traces the large-scale matter distribution through the IGM gas \cite{MiraldaEscude1996, Muecket1996, Theuns1998, Viel2004}. These approaches begin from a model of the primordial density fluctuations, evolve them forward under gravity, and predict the resulting \Lyaforest using the fluctuating Gunn--Peterson approximation (see Section~\ref{sec:simulating}). By iteratively comparing these predictions with the observed absorption field, they infer the underlying matter density distribution \cite{Horowitz2019, Porqueres2019, Horowitz2021, Horowitz2022_recons}. Unlike traditional tomographic techniques, which primarily interpolate between sparse sightlines, these methods explicitly incorporate a physical model of structure formation. This increased physical realism, however, comes at a substantial computational cost, since the reconstruction requires repeated forward modeling with gravity solvers.

To alleviate this limitation, recent work has employed variational autoencoders based on U-Net architectures to learn the inverse mapping between the \Lyaforest and the underlying matter density field directly from hydrodynamical simulations \cite{Maitra2025}. These machine-learning-based methods improve computational efficiency by orders of magnitude while matching, or even surpassing, the reconstruction fidelity of previous approaches. Although still at an early stage of development, such techniques represent promising alternatives to traditional tomographic methods.

\begin{figure}
    \centering
    \includegraphics[width=\linewidth]{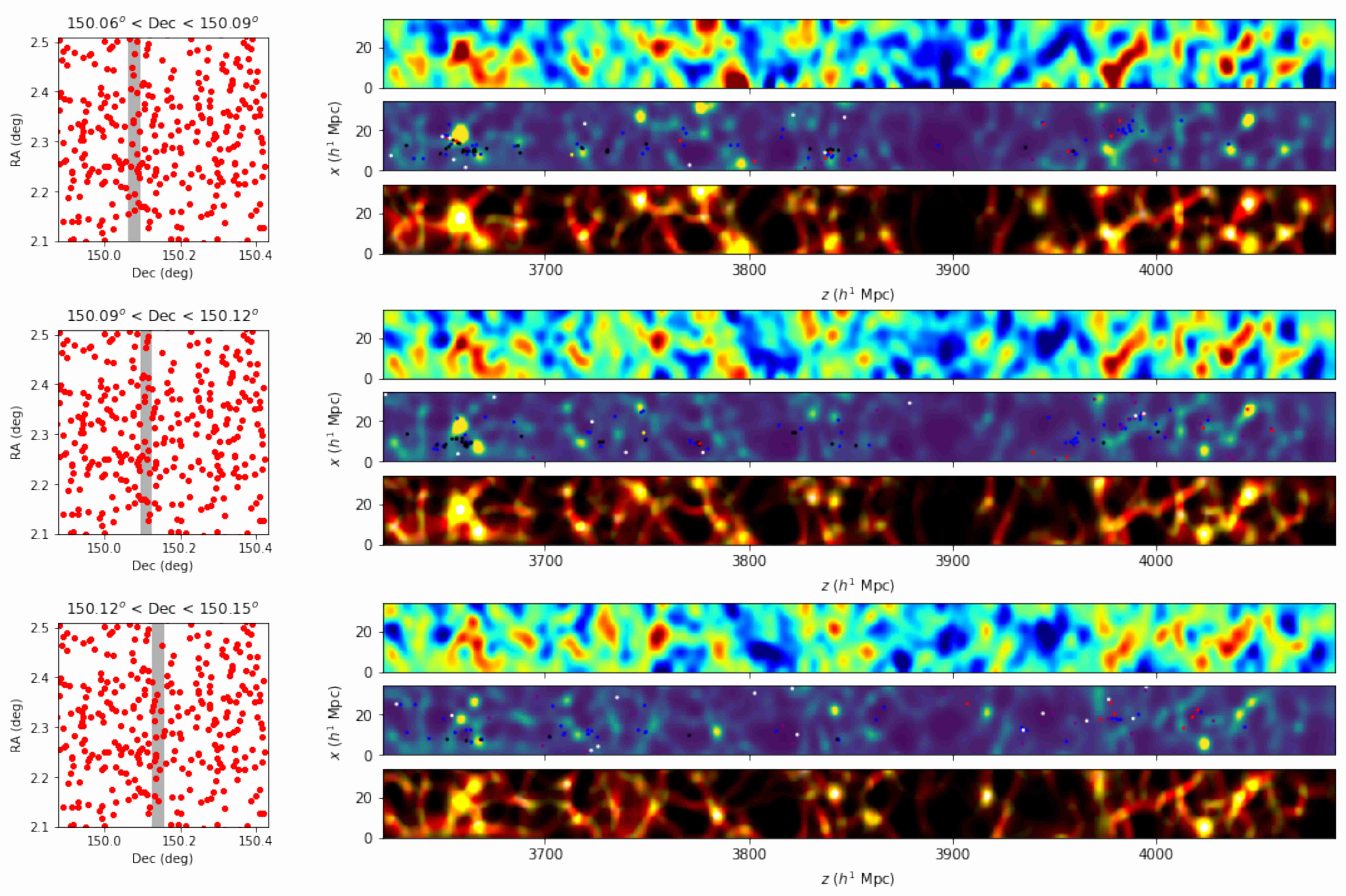}
    \caption{Tomographic reconstruction of large-scale structure from \Lyaforest observations in the CLAMATO survey \cite{Lee2018}. The left panel shows the angular distribution of quasars used to extract \Lyaforest spectra. In the right column, the top, middle, and bottom panels display the reconstructed IGM density field, the inferred dark matter distribution, and the resulting cosmic-web classification, respectively. Red regions correspond to overdensities, while blue regions indicate underdensities. The points overlaid in the middle panel mark the locations of galaxies within the surveyed field. Credit: Figure~14 from \cite{Horowitz2022_recons}.}
    \label{fig:recons}
\end{figure}


\section{Conclusions}
\label{sec:conclusions}

The \Lyaforest has become one of the most powerful probes of the distant Universe, providing unique constraints on the thermal and ionization state of the intergalactic medium, the growth of cosmic structure, the expansion history of the Universe, the nature of dark matter, and the physics of cosmic reionization. At the same time, fully exploiting the constraining power of the \Lyaforest has long been limited by the computational cost of the hydrodynamical simulations required to model the signal, as well as by the practical challenges involved in recovering the transmitted-flux field from noisy and contaminated quasar spectra. In this context, machine-learning (ML) techniques are no longer confined to isolated applications, but are increasingly being integrated throughout the full analysis pipeline, often providing substantial gains in computational efficiency, scalability, and information extraction.

\begin{itemize}

    \item {\bf Data reduction.} Convolutional neural networks (CNNs) have been used to automate the detection and characterization of \Lyaforest absorbers \cite{Cheng2022, Jalan2024}, high column density systems \cite{Parks2018, Wang2022_DLA, Chabanier2022, Brodzeller2025}, and quasars exhibiting broad absorption line features \cite{Busca2018, Guo2019, Rastegarnia2022, Moradi2024, Li2025}, improving the robustness and scalability of these analyses. In addition, CNNs, autoencoders, and related architectures have been applied to reconstruct the intrinsic quasar continuum at intermediate \cite{Fathivavsari2020, Bosman2021, Liu2021, Sun2023, Turner2024, Pistis2025, Hahn2025} and high redshift \cite{Durovcikova2020, Reiman2020, Durovcikova2024, Greig2024}, enabling access to larger-scale modes in future clustering analyses.

    \item {\bf Accelerating simulations.} U-Net architectures and conditional variational autoencoders have been used to generate high-fidelity \Lyaforest realizations from gravity-only simulations \cite{Harrington2022, Horowitz2022, Boonkongkird2023}, while generative adversarial networks have been employed to enhance the resolution of low-fidelity simulations \cite{Jacobus2023, Jacobus2025, Hafezianzadeh2025} or to generate \Lyaforest mocks directly without explicit simulation inputs \cite{ZamudioFernandez2019}.

    \item {\bf Accelerating likelihood inference.} Emulators --- fast surrogate models ranging from Gaussian processes \cite{Bird2019, Pedersen2021, Pedersen2023, Fernandez2022, Bird2023} to feed-forward neural networks \cite{CabayolGarcia2023, Molaro2023} and conditional normalizing flows \cite{ChavesMontero2025} --- have a long history of applications in cosmological and astrophysical analyses of the \Lyaforest. These include constraining dark matter properties \cite{Rogers2021a, Rogers2021b, Rogers2022, GarciaGallego2025}, the thermal and ionization state of the intergalactic medium \cite{Rorai2013, Rorai2017, Hiss2019, Walther2019, Jin2025, GonzalezHernandez2025a}, the topology of reionization \cite{Molaro2023}, and the small-scale matter power spectrum \cite{Fernandez2024, Walther2025, Ho2025, ChavesMontero2026}. Since these emulators are trained on computationally expensive hydrodynamical simulations, recent work has explored adaptive sampling strategies \cite{Rogers2019, Walther2021} and multi-fidelity approaches \cite{Fernandez2022, Bird2023, Fernandez2024, Ho2025} to alleviate the associated computational cost.

     \item {\bf Simulation-based inference.} Masked autoregressive flows have been employed to learn the likelihood function and infer IGM properties from Voigt-profile decompositions of the \Lyaforest \cite{Hu2022, Hu2024, Hu2025}, as well as to learn posterior distributions and constrain cosmological parameters \cite{Sinigaglia2026}. In parallel, ResNet architectures have been used to learn the likelihood-to-evidence ratio and characterize the epoch of reionization \cite{GonzalezHernandez2025b}. Furthermore, several studies have explored ML-based compression schemes designed to recover information beyond that captured by conventional summary statistics \cite{Charnock2018, Maitra2024}, while others directly employ ResNet architectures to extract information from spectral features and constrain the temperature--density relation of the IGM \cite{Nayak2024, Nayak2025}, yielding substantially tighter constraints than traditional summary-statistic-based methods \cite{Chang2025}.
    
    \item {\bf Field-level approaches.} CNNs have enabled pixel-level inference of the \Lya optical-depth field from noisy and saturated transmitted-flux data \cite{Huang2021}, as well as direct constraints on the thermal state of the IGM from quasar spectra \cite{Wang2022, Chen2023, Nasir2024}. These methods have also been employed to reconstruct the three-dimensional distribution of intergalactic gas --- including its density, temperature, and peculiar velocity fields --- from sparse \Lyaforest observations \cite{Ding2024}, as well as the underlying distribution of dark matter \cite{Horowitz2019, Porqueres2019, Horowitz2021, Horowitz2022_recons, Maitra2025}.
\end{itemize}

This transition is particularly important in the context of next-generation surveys. DESI is already delivering an unprecedented number of \Lyaforest spectra and JWST pushing \Lyaforest towards very high redshift, and future facilities such as DESI-II \cite{Schlegel2022}, the Stage-5 Spectroscopic Experiment (Spec-S5 \cite{Besuner2025}), the Wide Field Spectroscopic Telescope (WST \cite{Mainieri2024}), and the Extremely Large Telescope (ELT \cite{Japelj2019}) will increase the volume and precision of the available data. Fully exploiting the constraining power of these datasets will require inference pipelines capable of handling increasingly complex forward models, non-Gaussian observables, and high-dimensional models. In this context, ML-based emulators, simulation-based inference techniques, and field-level approaches are likely to become central components of future \Lyaforest analyses. At the same time, these methods rely heavily on the realism and parameter-space coverage of the simulations used for training, making robustness to modeling uncertainties and simulation systematics a critical challenge.
\section*{Acknowledgments}

I thank Marco Gatti and the members of the IFAE \Lya team for insightful comments and suggestions. I acknowledge financial support from the Spanish Ministry of Science and Innovation through the PID2024-159420NB-C41 project and the ``Excelencia Severo Ochoa'' program (CEX2024-001441-S from MICIU AEI 10.13039/501100011033) and the European Union through the ERC Consolidator Grant program (COSMO-LYA, grant agreement 101044612). Views and opinions expressed are however those of the authors only and do not necessarily reflect those of the European Union or the European Research Council Executive Agency. Neither the European Union nor the granting authority can be held responsible for them. IFAE is partially funded by the CERCA program of the Generalitat de Catalunya.

\bibliographystyle{JHEP.bst}
\bibliography{biblio}

\end{document}